\documentclass[aps,prd,superscriptaddress]{revtex4-2}

\usepackage{amsmath}
\usepackage{amssymb}
\usepackage{graphicx}
\usepackage{bm}

\begin{document}

\title{Collisionless damping of the gravitational instability in fuzzy dark matter: spectral shape and quantum-to-thermal crossover}

\author{Yosuke Matsumoto}
\email{ymatumot@chiba-u.jp}
\affiliation{Institute for Advanced Academic Research, Chiba University, Japan}

\author{Kohji Yoshikawa}
\affiliation{Center for Computational Sciences, University of Tsukuba, Japan}

\author{Naoki Yoshida}
\affiliation{Department of Physics, The University of Tokyo, Japan}

\date{\today}

\begin{abstract}
We present a quantum-kinetic linear theory of the gravitational instability in the context of fuzzy dark matter universe. Starting from the Wigner transport equation, we apply Landau's approach to the linearized Wigner--Poisson system and derive a kinetic dispersion relation that incorporates quantum effects exactly by introducing the plasma dispersion function. The growth rate as a function of wavenumber is characterized by a dimensionless quantum-to-thermal ratio $\alpha = k_{\mathrm{qJ}}/k_{\mathrm J}$, where $k_{\mathrm{qJ}}$ and $k_{\mathrm J}$ represent the quantum and thermal Jeans wavenumbers, respectively. We derive an analytic expression for the spectral slope at the cutoff wavenumber, revealing that the spectral shape undergoes a sharp transition across $\alpha \sim 0.5$. This implies a crossover from a thermally dominated kinetic regime, in which collisionless damping occurs via phase mixing and Landau resonance, to a regime dominated by quantum pressure. By applying these results to fuzzy dark matter, we show that the cutoff scale and its spectral shape depend sensitively on both the particle mass and the initial velocity dispersion, suggesting a method for simultaneously constraining these parameters through observations of the matter power spectrum. This framework provides a theoretical basis for future studies on the transition from early-phase thermal states to the formation of Bose-Einstein condensates in galactic structures.
\end{abstract}

\maketitle

\section{Introduction}
\label{sec:intro}
A collisionless system is composed of particles that have negligible direct interactions with one another. These particles equilibrate through interactions with fields, such as electromagnetic and gravitational fields. Understanding how energy is distributed in such systems has been a central topic. Collisionless damping, also referred to as Landau damping \citep{landauVIBRATIONSELECTRONICPLASMA1965}, is a key mechanism in this context and provides deep insights into how collisionless particles exchange energy with waves in a resonant manner. Specifically, electrons can be heated via Landau damping of plasma waves, and this type of electron heating has been observationally confirmed in fusion and space plasmas \citep{idaDirectObservationMassdependent2022,chenEvidenceElectronLandau2019}. Since the shape of the velocity distribution function is essential in Landau damping, the mechanism can be fully understood only through the kinetic treatment of the Vlasov (collisionless Boltzmann) equation.

A collisionless system also applies to the formation of cosmological structures, in which dark matter particles interact via a long-range self-gravitational potential. Although specific details about the particles, including their masses and velocity dispersions, remain poorly understood, the cold dark matter (CDM) model precisely explains the hierarchical structure of the universe at cosmological scales. Since the CDM model is generally treated as a zero-temperature model, initial density perturbations at all scales grow via the gravitational (Jeans) instability, forming scale-free structures in the early universe. Cosmological N-body simulations have shown the cosmic web evolution \citep{springelSimulationsFormationEvolution2005}, and the density structure at galaxy scales exhibits a cusp profile toward the center, which is known as the Navarro–Frenk–White (NFW) profile \citep{navarroStructureColdDark1996}. 

Observations of dwarf galaxies, however, have revealed a flat, cored density profile in the inner region, in contrast to the cuspy NFW profile predicted by dark-matter-only N-body simulations. More recently, this tension has been widely recognized as the diversity problem: dwarf galaxies with similar maximum circular velocities exhibit a much wider variety of inner dark matter slopes. This range includes both large-core profiles and profiles that are even steeper than the NFW profile. Reproducing this scatter in the slope of the inner density profile remains challenging even with hydrodynamic simulations that include baryonic feedback processes \citep{ohHIGHRESOLUTIONMASSMODELS2015, Oman2015, Hayashi2020, Sales2022}. A related issue known as the too-big-to-fail problem arises because the densest CDM subhalos are too centrally concentrated to host the observed Milky Way satellites. Whether these tensions between observations and the CDM model reflect new dark matter models or a poor understanding of baryonic feedback processes remains a matter of debate. Nonetheless, these persistent discrepancies at galactic scales continue to motivate alternative dark matter models beyond the standard CDM model. 

Among the proposed dark matter models, ultra-light dark matter (termed fuzzy dark matter or FDM) has attracted much attention for its potential to address the core-cusp problem \citep{huFuzzyColdDarkMatter2000,huiUltralightScalarsCosmological2017,huiWaveDarkMatter2021,ohHIGHRESOLUTIONMASSMODELS2015}. The mass of FDM is predicted to be extremely light, which results in a de Broglie wavelength comparable to galactic scales. Consequently, FDM is treated as a wave rather than as a collection of particles. Its dynamics can be described by the Schr\"odinger-Poisson (SP) equations or their equivalent Madelung form as the quantum hydrodynamics (QHD) equations \citep{madelungQuantentheorieHydrodynamischerForm1927,chavanisMassradiusRelationNewtonian2011a}. Since a solution to the hydrostatic equilibrium of the QHD equations represents a soliton-like density profile \citep{chavanisMassradiusRelationNewtonian2011a,chavanisMassradiusRelationNewtonian2011b}, and a similar density profile has been independently proposed by large-scale cosmological simulations based on the SP equations \citep{schiveCosmicStructureQuantum2014,moczSchrodingerPoissonVlasovPoissonCorrespondence2018,moczFirstStarFormingStructures2019}, the FDM model is emerging as a viable alternative to the standard CDM model.

FDM particles are typically regarded as bosons with many particles occupying the ground state as a Bose-Einstein condensate (BEC). Therefore, the state of FDM is generally represented by a single wave function. Although the medium is pressureless, the linearized QHD equations indicate that a quantum pressure can resist gravitational collapse. This results in a sharp drop in the growth rate of gravitational instability in the very vicinity of the quantum Jeans wavenumber \citep{chavanisMassradiusRelationNewtonian2011a}.

The mass of FDM can be estimated using the mass-radius relation of a soliton core. By relating the typical total mass of the soliton core to the quantum Jeans mass, a mass of $\sim 3 \times 10^{-22}~\text{eV}$ can be obtained \citep{chavanisMassradiusRelationNewtonian2011a}. This value, however, contradicts the constraint on mass from the Lyman-alpha forest observations, which have suggested a mass greater than $2 \times 10^{-21}~\text{eV}$ \citep{armengaudConstrainingMassLight2017,huiUltralightScalarsCosmological2017,irsicFirstConstraintsFuzzy2017}. This discrepancy remains a tension regarding the FDM mass.

In this paper, we consider a scenario in which the FDM particles are initially in incoherent states and transition into the BEC state in gravitationally collapsing regions after the gravitational instability saturates \citep{levkovGravitationalBoseEinsteinCondensation2018,barorRelaxationFuzzyDark2021,liuInterferenceGravitationalInstability2025} during the early phase of structure formation. We first present a kinetic linear theory of the gravitational instability that fully incorporates the quantum effects, and then discuss how a finite velocity dispersion affects the growth of the instability, with particular focus on the spectral shape around the cutoff wavenumber. The paper is organized as follows. In Section~\ref{sec:base}, we introduce the Wigner distribution function and derive the Wigner transport equation, which governs the time evolution of the quantum phase-space distribution. In Section~\ref{sec:linear}, we apply Landau's approach to the linearized Wigner--Poisson system to derive the full quantum-kinetic dispersion relation of the gravitational instability. In Section~\ref{sec:stability}, we solve the dispersion relation numerically and derive analytic expressions for the cutoff wavenumber and the spectral slope at the cutoff. In Section~\ref{sec:implication}, we quantify these results in physical units relevant to FDM and discuss their implications for observations of the matter power spectrum. Section~\ref{sec:conclusion} summarizes our conclusions.

\section{Wigner transport equation}
\label{sec:base}
When particles can exist in different states, each with an associated probability, it is more convenient to work with an ensemble-averaged density matrix instead of the individual wave functions. The density matrix is expressed as
\begin{equation}
    \hat \rho = \sum_j p_j |\phi_j\rangle \langle \phi_j |,
\end{equation}
where $|\phi_j\rangle$ represents the wave function of a given state and $p_j$ is the corresponding probability of that state. It can be shown that the density matrix satisfies the von Neumann equation
\begin{equation}
  i\hbar\frac{\partial \hat\rho}{\partial t} = \hat H \hat\rho - \hat\rho\hat H = \left[\hat H, \hat\rho\right],
  \label{eq:vonNeumann}
\end{equation}
where $\hbar$ is the reduced Planck constant and $\hat H$ is the Hamiltonian operator.

For the linear stability analysis, it is mathematically useful to map the density matrix onto classical phase space spanned by the coordinate $x$ and the momentum $p$ using the Wigner transform
\begin{equation}
\begin{aligned}
  f_\mathrm{w}(x,p) &= \frac{1}{2\pi\hbar} \int_{-\infty}^\infty \left\langle x+\frac{\xi}{2}\right|\hat\rho\left|x-\frac{\xi}{2}\right\rangle\exp{\left(-i\frac{p}{\hbar} \xi\right)}~d\xi \\
   &= \frac{1}{2\pi\hbar} \int_{-\infty}^\infty \rho\left(x+\frac{\xi}{2},x-\frac{\xi}{2}\right)\exp{\left(-i\frac{p}{\hbar} \xi\right)}~d\xi,
\end{aligned}
\label{eq:wigner_f}
\end{equation}
where the system is assumed to be homogeneous and isotropic, and we reduce the analysis to one spatial direction and one momentum direction without loss of generality. This expression represents the Wigner distribution function \citep{wignerQuantumCorrectionThermodynamic1932}, which is a quasi-probability distribution in phase space. To obtain the time evolution of the Wigner distribution function, we also apply the Wigner transform to the right-hand side of Eq. \eqref{eq:vonNeumann}. After performing some algebraic manipulations, we arrive at the Wigner transport equation \citep{moyalQuantumMechanicsStatistical1949}
\begin{equation}
  \frac{\partial f_\mathrm{w}}{\partial t} = -\frac{p}{m}\frac{\partial f_\mathrm{w}}{\partial x}
 +\frac{1}{i\hbar}\frac{1}{2\pi\hbar}\int_{-\infty}^\infty\left(V\left(x+\frac{\xi}{2}\right)-V\left(x-\frac{\xi}{2}\right)\right)\rho\left(x+\frac{\xi}{2},x-\frac{\xi}{2}\right)\exp{\left(-i\frac{p}{\hbar} \xi\right)}~d\xi,
\label{eq:wigner_transport}
\end{equation}
where $ V(x) $ represents the potential energy, with which the classical Hamiltonian is expressed as $H=\frac{p^2}{2m}+V(x)$ for a particle with a mass $m$. If we express $V(x)$ in terms of its Fourier amplitude $\tilde V(k)$, Eq. \eqref{eq:wigner_transport} can be rewritten as 
\begin{equation}
  \frac{\partial f_\mathrm{w}}{\partial t} = -\frac{p}{m}\frac{\partial f_\mathrm{w}}{\partial x}  + \frac{1}{i\hbar}\frac{1}{2\pi}\int_{-\infty}^\infty dk\ \tilde V(k)\left[f_\mathrm{w}\left(x,p-\frac{\hbar k}{2}\right)-f_\mathrm{w}\left(x,p+\frac{\hbar k}{2}\right)\right] \exp{(ikx)},
  \label{eq:wigner_transport_fourier}
\end{equation}
which was found to be a suitable form for the following linear analysis \citep{mendoncaWavekineticApproachSchrodinger2019,mendoncaLandauDampingParticle2023}.

The Wigner transport equation (Eq.~\eqref{eq:wigner_transport_fourier}) can also be written in an equivalent differential form, whose semi-classical expansion recovers the Vlasov equation with quantum corrections \citep{moyalQuantumMechanicsStatistical1949,bertrandClassicalVlasovPlasma1980,koppSolvingVlasovEquation2017,moczSchrodingerPoissonVlasovPoissonCorrespondence2018,gomesQuantumKineticTheory2023} (See also Appendix~\ref{sec:appendix_semiclassical}). For the linear analysis below, we use the form of Eq.~\eqref{eq:wigner_transport_fourier}.

\section{Quantum-kinetic linear theory}
\label{sec:linear}
\subsection{Linearized Wigner--Poisson system}
We examine the Wigner--Poisson system with the Hamiltonian $H = \frac{p^2}{2m} + m\Phi(x)$, in which the gravitational potential $\Phi$ satisfies the Poisson equation
\begin{equation}
  \frac{\partial^2 \Phi}{\partial x^2} = 4\pi G \rho = 4\pi G m \int_{-\infty}^{+\infty} f_\mathrm{w}~dp,
\end{equation}
where $G$ is the gravitational constant.

We adopt Landau's approach used for the kinetic plasma wave to derive the dispersion relation of the gravitational instability. We decompose $f_\mathrm{w} = f_\mathrm{0w}(v) + \delta f_\mathrm{w}(t,x,v)$ and $\Phi = \delta\Phi(t,x)$ with the equilibrium gravitational potential $\Phi_0 = 0$ by subtracting the average mass density in the Poisson equation. The so-called Jeans swindle can be mathematically justified by considering the effects of an expanding universe \citep{huiWaveDarkMatter2021}. We write the linearized equations as
\begin{equation}
  \frac{\partial \delta f_\mathrm{w}}{\partial t} 
  = -v\frac{\partial \delta f_\mathrm{w}}{\partial x} 
    + \frac{m}{i\hbar}\frac{1}{2\pi} \int_{-\infty}^\infty \delta\tilde\Phi(k)\left[f_\mathrm{0w}\left(v-\frac{\hbar k}{2m}\right)-f_\mathrm{0w}\left(v+\frac{\hbar k}{2m}\right)\right]\exp{(ikx)}~dk,
\end{equation}
\begin{equation}
    \nabla^2 \delta \Phi = 4\pi G m \int_{-\infty}^{+\infty} \delta f_\mathrm{w}\ dv,
\end{equation}
where $v = p/m$. Applying the Laplace transform for time with frequency $s = \gamma - i\omega$ and the Fourier transform for space with wavenumber $k$, we obtain an equation for $\delta \Phi(s,k)$ as
\begin{equation}
  \delta\Phi(s,k) = i\frac{\displaystyle \frac{4\pi Gm}{k^3}\int_{-\infty}^{+\infty}\frac{\delta f_\mathrm{w}(0,k,v)}{v-is/k}~dv}
  {\displaystyle 1-\frac{4\pi Gm^2}{\hbar k^3}\int_{-\infty}^{+\infty}\frac{f_\mathrm{0w}(v-\frac{\hbar k}{2m})-f_\mathrm{0w}(v+\frac{\hbar k}{2m})}{v-is/k}~dv},
\end{equation}
where $\delta f_\mathrm{w}(0,k,v)$ represents the initial value of the perturbed distribution function. The inverse Laplace transform gives its time evolution
\begin{equation}
  \delta\Phi(t,k) = \frac{1}{2\pi i}\int_{\sigma-i\infty}^{\sigma+i\infty}\delta\Phi(s,k)\exp{(st)}\ ds.
\end{equation}
The integral is evaluated along the line $\Re(s) = \sigma$ aligned with the imaginary axis, which must be sufficiently large to ensure that all poles of $\delta \Phi(s,k)$ are located in the left half of the complex plane. The denominator (numerator) must be analytically continued in $\Re{(s)} \leq 0$, allowing the inverse Laplace transform to be evaluated based on the poles within the closed integral contour (Cauchy's residue theorem). After a long time, the pole in the exponent of the exponential function determines whether the mode $\delta\Phi(t,k)$ grows or decays. Consequently, the problem reduces to identifying the poles of $\delta\Phi(s,k)$, which appear for
\begin{equation}
1-\frac{4\pi Gm^2}{\hbar k^3} \int_{-\infty}^{+\infty} \frac{\displaystyle f_\mathrm{0w}\left(v-\frac{\hbar k}{2m}\right)-f_\mathrm{0w}\left(v+\frac{\hbar k}{2m}\right)}{v-is/k}~dv = 0.
\label{eq:qk_dispersion}
\end{equation}
This is the kinetic dispersion relation of the gravitational instability fully incorporating quantum effects \citep{mendoncaWavekineticApproachSchrodinger2019,barorRelaxationFuzzyDark2021}.

\subsection{Zero-temperature limit}
Before examining the full kinetic dispersion relation, we verify consistency in the zero-temperature limit by adopting $f_\mathrm{0w}(v) = n_0\,\delta(v)$. Eq. \eqref{eq:qk_dispersion} is simplified to 
\begin{equation}
  1-\frac{4\pi Gn_0m^2}{\hbar k^3}\left(\int_{-\infty}^{+\infty}\frac{\displaystyle\delta(v-\frac{\hbar k}{2m})-\delta(v+\frac{\hbar k}{2m})}{v-is/k}\ dv\right) 
  = 1-\frac{4\pi Gn_0m^2}{\hbar k^3}\left(\frac{1}{\displaystyle\frac{\hbar k}{2m}-is/k}+\frac{1}{\displaystyle\frac{\hbar k}{2m}+is/k}\right) = 0.
\end{equation}
We finally obtain 
\begin{equation}
  s = \sqrt{4\pi Gmn_0 - \frac{\hbar^2 k^4}{4m^2}}= \sqrt{4\pi G\rho_0\left(1-\frac{k^4}{k_{\mathrm{qJ}}^4}\right)},
  \label{eq:dispersion_qhd}
\end{equation}
where $\rho_0=mn_0$ is the background mass density, and the quantum Jeans wavenumber is defined as
\begin{equation}
  k_{\mathrm{qJ}} \equiv \left(\frac{16\pi Gm^2\rho_0}{\hbar^2}\right)^{1/4}.
  \label{eq:kqJ}
\end{equation}
The quantum-kinetic dispersion relation reproduces the one from the QHD equations \citep{chavanisMassradiusRelationNewtonian2011a} in the zero-temperature limit.

\subsection{Full quantum-kinetic dispersion relation}
A kinetic description of self-gravitating FDM has been developed by \citet{barorRelaxationFuzzyDark2021} in the context of the relaxation of FDM halos, in which the background $f_\mathrm{0w}$ was taken to be a Maxwell distribution,
\begin{equation}
  f_\mathrm{0w}(v) = \frac{n_0}{\sqrt{\pi}\,v_\mathrm{t}}\exp{\left(-\frac{v^2}{v_\mathrm{t}^2}\right)}.
\end{equation}
We also assume that the equilibrium is a spatially homogeneous, completely incoherent mixture of free-particle eigenstates, and $f_\mathrm{0w}$ can be simplified to a positive-definite classical Maxwell distribution, where $v_\mathrm{t}$ encodes the random phase differences among the superposed initial wavefunctions. Thus, our derived dispersion relation below is consistent with their dielectric function. We expand on their analysis by illustrating the complete growth rate profiles for various background quantum-to-thermal ratios, and examining the nature of collisionless damping by analyzing angular frequency spectra. Furthermore, we derive the spectral slope at the cutoff wavenumber to highlight implications for the tension in the FDM mass.

The integral in Eq. \eqref{eq:qk_dispersion} is written as
\begin{equation}
  \frac{n_0}{\sqrt{\pi}v_\mathrm{t}}\int_{-\infty}^{+\infty}
  \frac{\displaystyle\exp\left\{-\frac{\left(v-\frac{\hbar k}{2m}\right)^2}{v_\mathrm{t}^2}\right\}
       -\exp\left\{-\frac{\left(v+\frac{\hbar k}{2m}\right)^2}{v_\mathrm{t}^2}\right\}}
       {v-is/k} dv.
  \label{eq:integ_maxwell}
\end{equation}
Next, we normalize the velocity to $v_\mathrm{t}$ as $z = v/v_\mathrm{t}$, $dv = v_\mathrm{t}dz$, and $\zeta = is/(kv_\mathrm{t})$, and define the thermal de Broglie wavenumber $k_{\mathrm q} = mv_\mathrm{t}/\hbar$. Then Eq. \eqref{eq:integ_maxwell} is rewritten as
\begin{equation}
  \frac{n_0}{\sqrt{\pi}v_\mathrm{t}}
  \left(
    \int_{-\infty}^{+\infty}
    \frac{\exp\left\{-\left(z-\frac{k}{2k_{\mathrm q}}\right)^2\right\}}{z-\zeta}\ dz
    -
    \int_{-\infty}^{+\infty}
    \frac{\exp\left\{-\left(z+\frac{k}{2k_{\mathrm q}}\right)^2\right\}}{z-\zeta}\ dz
  \right).
  \label{eq:integ_maxwell2}
\end{equation}
Replacing the integration variable $z$ with $z\mp k/2k_{\mathrm q}$ in each integral, Eq. \eqref{eq:integ_maxwell2} is further rewritten as
\begin{equation}
  \frac{n_0}{\sqrt{\pi}v_\mathrm{t}}
  \left(
    \int_{-\infty}^{+\infty}
    \frac{\exp{(-z^2)}}{\displaystyle z-\left(\zeta-\frac{k}{2k_{\mathrm q}}\right)}\ dz
    -
    \int_{-\infty}^{+\infty}
    \frac{\exp{(-z^2)}}{\displaystyle z-\left(\zeta+\frac{k}{2k_{\mathrm q}}\right)}\ dz
  \right)
  = 
  \frac{n_0}{v_\mathrm{t}}
  \left(Z\left(\zeta-\frac{k}{2k_{\mathrm q}}\right)-Z\left(\zeta+\frac{k}{2k_{\mathrm q}}\right)\right)
  \label{eq:integ_maxwell3}
\end{equation}
where
\begin{equation}
  Z(\zeta) \equiv \frac{1}{\sqrt{\pi}}
  \int_{-\infty}^{+\infty}\frac{\exp{(-z^2)}}{z-\zeta}\ dz
  \label{eq:Z}
\end{equation}
is the plasma dispersion function \citep{friedPlasmaDispersionFunction1961,xieRapidComputationPlasma2024} (See also Appendix \ref{sec:appendix_plasma}). 

Substituting Eq. \eqref{eq:integ_maxwell3} into Eq. \eqref{eq:qk_dispersion}, we obtain \citep{barorRelaxationFuzzyDark2021}
\begin{equation}
  1 - \frac{k_{\mathrm J}^2k_{\mathrm q}}{2k^3}\left[Z\left(\zeta-\frac{k}{2k_{\mathrm q}}\right)-Z\left(\zeta+\frac{k}{2k_{\mathrm q}}\right)\right] = 0,
  \label{eq:qkz}
\end{equation}
where we have defined the Jeans wavenumber
\begin{equation}
    k_{\mathrm J} \equiv \sqrt{\frac{8\pi G\rho_0}{v_\mathrm{t}^2}}.
\end{equation}

\subsection{Classical limit}
In the classical limit of $\hbar \to 0$ ($k_{\mathrm q}\to\infty$), we can expand $Z(\zeta\pm k/(2k_{\mathrm q}))$ in a Taylor series around $\zeta$ as
\begin{equation}
  Z\left(\zeta\pm\frac{k}{2k_{\mathrm q}}\right) = Z(\zeta) \pm Z^\prime(\zeta)\frac{k}{2k_{\mathrm q}}+\frac{Z^{\prime\prime}(\zeta)}{2}\left(\frac{k}{2k_{\mathrm q}}\right)^2 \pm \cdots
\end{equation}
The difference between the plasma dispersion functions in Eq. \eqref{eq:qkz} is
\begin{equation}
  Z\left(\zeta-\frac{k}{2k_{\mathrm q}}\right) - Z\left(\zeta+\frac{k}{2k_{\mathrm q}}\right) = -Z^\prime(\zeta)\frac{k}{k_{\mathrm q}} + O\left(\hbar^3\right)
  = 2\left(1+\zeta Z(\zeta)\right)\frac{k}{k_{\mathrm q}} + O\left(\hbar^3\right),
  \label{eq:Zdiff}
\end{equation}
where we have used the first derivative of the plasma dispersion function (Eq. \eqref{eq:Zprime}). Substituting Eq. \eqref{eq:Zdiff} to Eq. \eqref{eq:qkz}, we obtain
\begin{equation}
  \lim_{\hbar \to 0} 1 - \frac{k_{\mathrm J}^2k_{\mathrm q}}{2k^3}\left[2\left(1+\zeta Z(\zeta)\right)\frac{k}{k_{\mathrm q}} + O\left(\hbar^3\right)\right]  = 1 - \frac{k_{\mathrm J}^2}{k^2}\left(1 + \zeta Z(\zeta)\right) = 0,
  \label{eq:disp_classical}
\end{equation}
which is exactly the classical kinetic dispersion relation \citep{binneyGalacticDynamicsSecond2011,yoshikawaDIRECTINTEGRATIONCOLLISIONLESS2013}.

\section{Stability analysis}
\label{sec:stability}
We solve Eq. \eqref{eq:qkz} numerically for $\zeta=is/(kv_\mathrm{t})=(i\gamma+\omega)/(kv_\mathrm{t})$. Thus, the imaginary part of $\zeta$ represents growth or damping, whereas the real part represents the oscillatory (propagating) modes. The plasma dispersion function can be expressed in relation to the Faddeeva function $w(\zeta)$ as
\begin{equation}
  Z(\zeta) = i\sqrt{\pi}w(\zeta),
\end{equation}
which is available as a special function \texttt{scipy.special.wofz} in the Python SciPy library \citep{virtanenSciPy10Fundamental2020}. We introduce a dimensionless parameter $\alpha = k_{\mathrm{qJ}}/k_{\mathrm J}$ to characterize the importance of the quantum to the thermal effects and analyze for different values of $\alpha$. 

\begin{figure}
\includegraphics[width=1.0\linewidth]{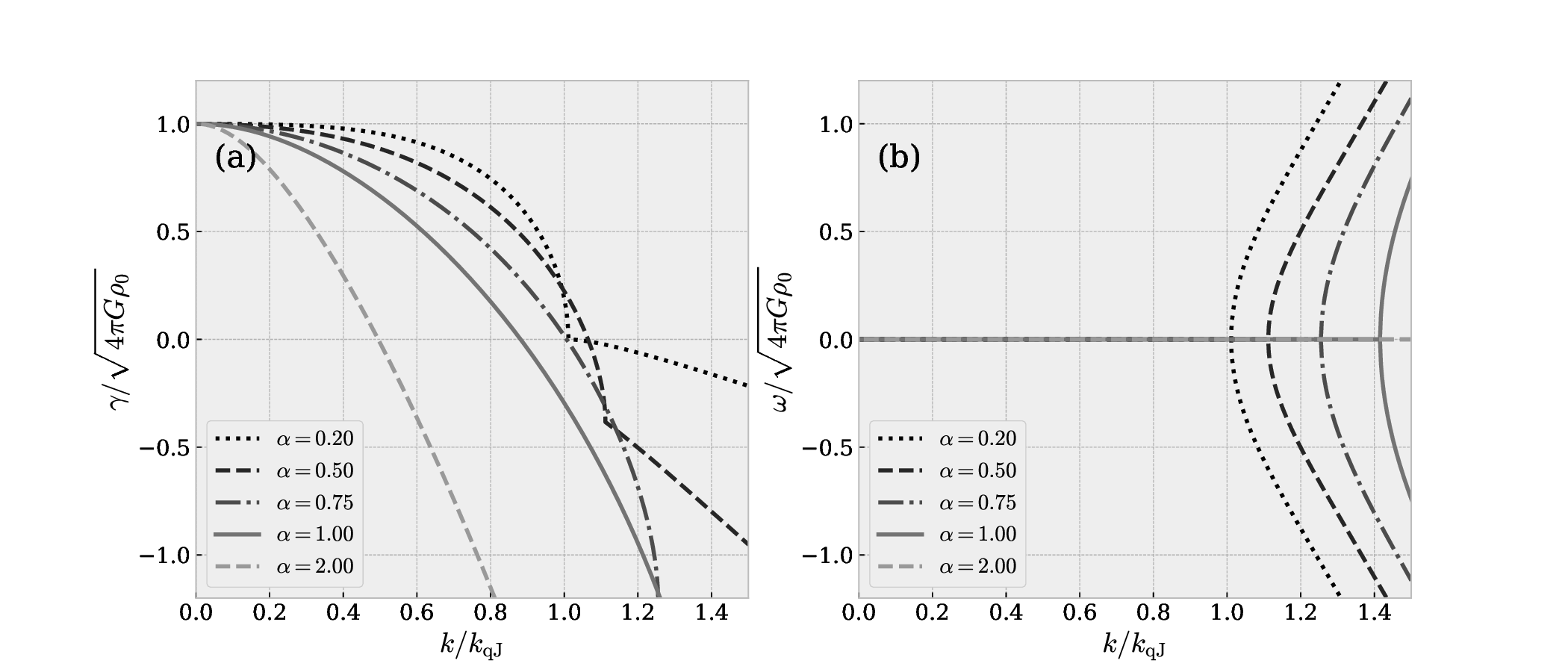}
\caption{Numerical solutions of the dispersion relation with various $\alpha$ for (a) the growth rate (imaginary part) and (b) the angular frequency (real part) as a function of the wavenumber $k$. The frequencies and the wavenumber are normalized to $\sqrt{4\pi G \rho_0}$ and $k_{\mathrm{qJ}}$, respectively.}
\label{omega-k}
\end{figure}

Figure \ref{omega-k} shows the results for $\alpha$ ranging from $0.2$ to $2.0$. The growth rate profiles (Fig. \ref{omega-k}(a)) transition from a classical kinetic profile for $\alpha = 2.0$ to an asymptotic QHD result for $\alpha = 0.2$ through intermediate profiles. The cutoff wavenumber $k_{\mathrm c}$, at which the system transitions from growing modes to damping modes, is characterized by the classical Jeans wavenumber $k_{\mathrm J}$ for large $\alpha$ and by the quantum Jeans scale $k_{\mathrm{qJ}}$ for small $\alpha$. However, the spectral shapes differ significantly within these profiles. In all scenarios, the modes are primarily unstable with zero frequency for wavenumbers below $\sim k_{\mathrm{qJ}}$. Oscillatory modes (Fig. \ref{omega-k}(b)) emerge in the regions where the wavenumber exceeds $k_{\mathrm{qJ}}$. However, these oscillatory modes are fundamentally damping modes, since their frequencies have negative imaginary parts as long as the velocity dispersion remains finite.

The damping rate in this region exhibits a distinct break, at which a real part of the frequency emerges. Since gravitational instability is essentially a zero-frequency mode, the damping at wavenumbers below the break is generally attributed to ballistic motion of particles in the high-energy tail of the distribution function, which smears out the density fluctuations (phase mixing). In the large-wavenumber region above the break, the quantum pressure introduces a counterforce, leading to the emergence of the real part of the frequency. In this regime, the gravitational mode is a propagating one with a phase speed at which density perturbations can be damped out via the Landau resonance. Therefore, the break represents a transition from non-resonant to resonant collisionless damping.

\subsection{Cutoff wavenumber}
It would be valuable to provide the cutoff wavenumber $k_{\mathrm c}$ for different values of $\alpha$ in relation to the smallest structures in the universe. By setting $\zeta = 0$ and using the property of the plasma dispersion function (Eq. \eqref{eq:Z_full}), we derive expressions for $Z(\pm k_{\mathrm c}/(2k_{\mathrm q}))$ as
\begin{equation}
  Z\left(\pm\frac{k_{\mathrm c}}{2k_{\mathrm q}}\right) = -2F\left(\pm\frac{k_{\mathrm c}}{2k_{\mathrm q}}\right) + i\sqrt{\pi}\exp{\left(-\frac{k_{\mathrm c}^2}{4k_{\mathrm q}^2}\right)}.
\end{equation}
Substituting these expressions into the dispersion relation (Eq. \eqref{eq:qkz}), we obtain 
\begin{equation}
  1 + \frac{k_{\mathrm J}^2k_{\mathrm q}}{k_{\mathrm c}^3} \left[F\left(-\frac{k_{\mathrm c}}{2k_{\mathrm q}}\right) - F\left(\frac{k_{\mathrm c}}{2k_{\mathrm q}}\right)\right] = 0.
\end{equation}
Since $F(-u) = -F(u)$,
\begin{equation}
  F\left(\frac{k_{\mathrm c}}{2k_{\mathrm q}}\right) = \frac{k_{\mathrm c}^3}{2k_{\mathrm J}^2k_{\mathrm q}},
  \label{eq:cutoff_exact}
\end{equation}
which can be solved numerically for $k_{\mathrm c}$.

For $k_{\mathrm c}\ll k_{\mathrm q}$, we can use the Taylor expansion of $F(u)\approx u - \frac{2}{3}u^3$. This leads to
\begin{equation}
  \frac{k_{\mathrm c}}{2k_{\mathrm q}} - \frac{1}{3}\frac{k_{\mathrm c}^3}{4k_{\mathrm q}^3}
  = \frac{k_{\mathrm c}^3}{2k_{\mathrm J}^2k_{\mathrm q}},
\end{equation}
giving an approximate closed-form solution of
\begin{equation}
  k_{\mathrm c} \approx \sqrt{\frac{6}{6\frac{k_{\mathrm q}^2}{k_{\mathrm J}^2}+1}}k_{\mathrm q} = \sqrt{\frac{3\alpha^2}{3\alpha^4+1}}k_{\mathrm{qJ}}.
  \label{eq:cutoff_approx}
\end{equation}
For $\alpha \gg 1$, $k_{\mathrm c}$ asymptotically approaches $ k_{\mathrm{qJ}}/\alpha = k_{\mathrm J}$.
\begin{figure}
\includegraphics[width=1.0\linewidth]{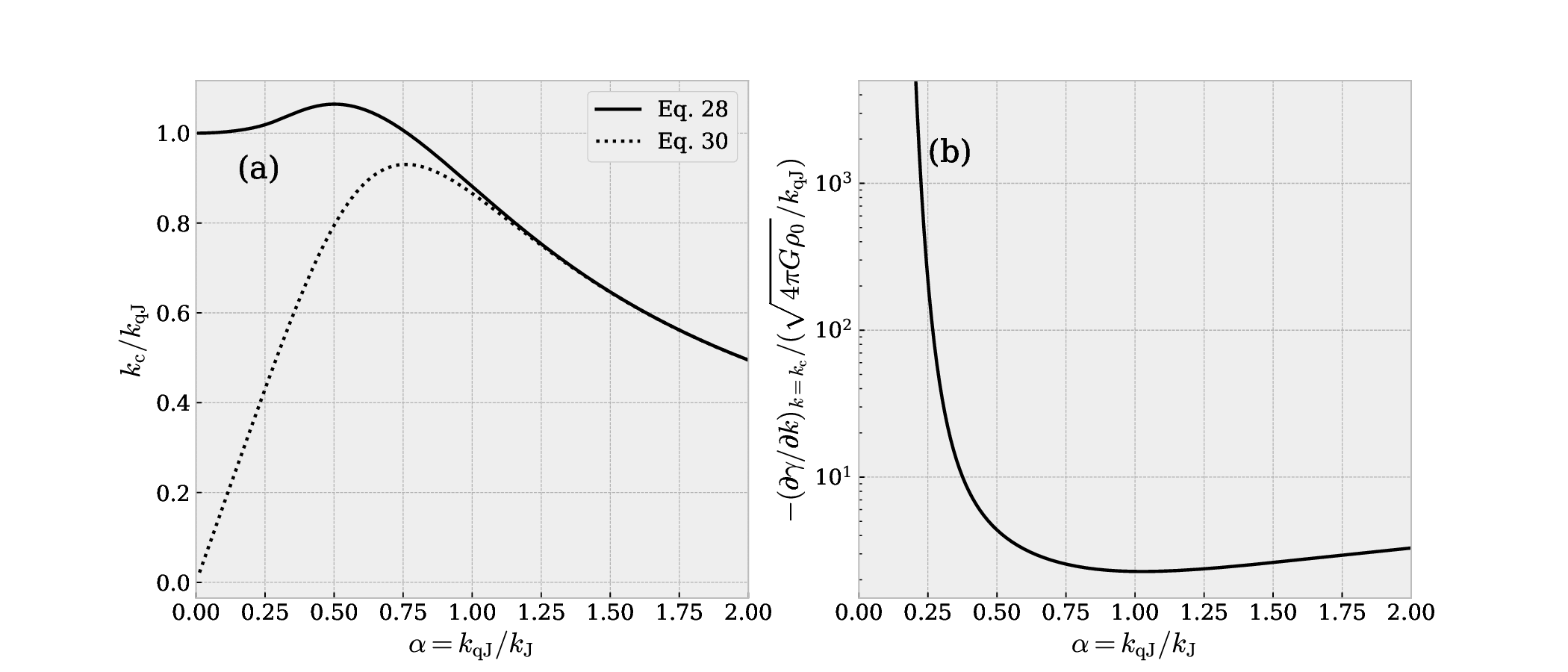}
\caption{(a) Numerical solutions to Eq. \eqref{eq:cutoff_exact} (solid line) and the approximate solution of Eq. \eqref{eq:cutoff_approx} (dotted line) for the cutoff wavenumber normalized to $k_{\mathrm{qJ}}$ for various values of $\alpha$. (b) Spectral slope at $k=k_{\mathrm c}$ normalized to $\sqrt{4\pi \rho_0 G}/k_{\mathrm{qJ}}$ for $\alpha$. A negative sign is applied to the slope to clarify the visual representation on a logarithmic scale.}
\label{fig:kcutoff-alpha}
\end{figure}

Figure \ref{fig:kcutoff-alpha} (a) exhibits the numerical solution to Eq. \eqref{eq:cutoff_exact} (solid line) alongside the approximate solution of Eq. \eqref{eq:cutoff_approx} (dotted line). The closed-form solution accurately predicts the cutoff wavenumber only for $\alpha > 1$. For $\alpha <1$, the numerical solution shows that the cutoff wavenumber varies around $k_{\mathrm{qJ}}$, reaching a peak value of $k_{\mathrm c}/k_{\mathrm{qJ}} \sim 1.06$ at $\alpha \sim 0.5$. It is noteworthy that the condition $k_{\mathrm c} = k_{\mathrm{qJ}}$ is met at two distinct values of $\alpha = 0$ and $\alpha \sim 0.76$. The behavior of the cutoff wavenumber between the classical and quantum Jeans scales is consistent with the critical wavenumber reported by \citet{barorRelaxationFuzzyDark2021}. We additionally resolve its non-monotonic variation for $\alpha \lesssim 0.76$, and examine the spectral slope around it in the next subsection.

The asymptotic behavior toward $\alpha = 0$ establishes a direct correspondence with a soliton core in the QHD picture. In the QHD model, a soliton forms at the length scale within which quantum pressure balances gravity, yielding a core radius of $r_{\mathrm c} \sim 1/k_{\mathrm{qJ}}$ \citep{chavanisMassradiusRelationNewtonian2011a}. The kinetic theory recovers this in the zero-temperature limit, confirming physical consistency. For $\alpha \lesssim 0.76$, the kinetic and QHD scales remain close, so the nonlinear structures expected to form from modes at $k \lesssim k_{\mathrm c}$ have a characteristic size comparable to the QHD soliton. For $\alpha \gtrsim 0.76$, the kinetic cutoff shifts to wavenumbers below $k_{\mathrm{qJ}}$, so velocity dispersion suppresses structure at scales larger than the soliton core. This separation of the kinetic and QHD scales, and its dependence on $\alpha$, underlies the relation between the soliton (QHD) scale and the cutoff of the matter power spectrum, which we quantify in Section~\ref{sec:implication}.

\subsection{Spectral slope}
While the cutoff wavenumber varies around $k_{\mathrm{qJ}}$ as long as $\alpha \lesssim 0.76$ as indicated by Figure \ref{fig:kcutoff-alpha}(a), the spectral slope around the cutoff wavenumber differs significantly across the same range of $\alpha$. The result from the QHD model (Eq. \eqref{eq:dispersion_qhd}) indicates that the gradient of the growth rate curve with respect to $k$ at $k_{\mathrm{qJ}}$ is
\begin{equation}
\left (\frac{\partial \gamma}{\partial k}\right )_{k=k_{\mathrm{qJ}}} = -\left(\frac{8 \pi G \rho_0\frac{k^3}{k_{\mathrm{qJ}}^4}}{\sqrt{4 \pi G \rho_0\left( 1 - \frac{k^4}{k_{\mathrm{qJ}}^4}\right)}}\right)_{k=k_{\mathrm{qJ}}} = -\infty.
\end{equation}
In contrast, Figure \ref{omega-k} (a) indicates that the spectral shapes with $\alpha > 1$ at the cutoff wavenumber show smoother profiles reflecting the collisionless damping. This change in the spectral shape suggests a transition from the classical kinetic profiles to those resembling the QHD characteristic. To quantify this transition, we derive $\left(\partial \gamma/\partial k\right)_{k=k_{\mathrm c}}$ from the dispersion relation (Eq. \eqref{eq:qkz}). From the implicit function theorem, 
\begin{equation}
\frac{\partial \gamma}{\partial k} = - \frac{\frac{\partial D(\gamma,k)}{\partial k}}{\frac{\partial D(\gamma,k)}{\partial \gamma}},
\end{equation}
where $D(\gamma,k)=0$ stands for the dispersion relation.

First, we consider the numerator as
\begin{align}
\frac{\partial D(\gamma,k)}{\partial k} = \frac{3k_{\mathrm J}^2 k_{\mathrm q}}{2k^4}\left( Z\left(\zeta-\frac{k}{2k_{\mathrm q}}\right) 
 - Z\left(\zeta+\frac{k}{2k_{\mathrm q}}\right)\right) \nonumber \\
 -\frac{k_{\mathrm J}^2k_{\mathrm q}}{2k^3}\left[\frac{2}{k_{\mathrm q}}+\left(\frac{2\zeta}{k}+\frac{1}{k_{\mathrm q}}\right)\left(\zeta -\frac{k}{2k_{\mathrm q}}\right)Z\left(\zeta - \frac{k}{2k_{\mathrm q}}\right) \right . \nonumber \\
 \left . -\left(\frac{2\zeta}{k}-\frac{1}{k_{\mathrm q}}\right)\left(\zeta + \frac{k}{2k_{\mathrm q}}\right)Z\left(\zeta + \frac{k}{2k_{\mathrm q}}\right)\right],
\end{align}
where we have used Eq. \eqref{eq:Zprime} for $Z^\prime$. At the cutoff wavenumber where $\zeta=0$, this reduces to
\begin{equation}
\left(\frac{\partial D(\gamma,k)}{\partial k}\right)_{k=k_{\mathrm c}} =\frac{6k_{\mathrm J}^2 k_{\mathrm q}}{k_{\mathrm c}^4}F\left(\frac{k_{\mathrm c}}{2k_{\mathrm q}}\right)-\frac{k_{\mathrm J}^2}{k_{\mathrm c}^3}\left[1 - \frac{k_{\mathrm c}}{k_{\mathrm q}} F\left(\frac{k_{\mathrm c}}{2k_{\mathrm q}}\right)\right].
\end{equation}
Using the relation at the cutoff wavenumber $F(k_{\mathrm c}/2k_{\mathrm q})=k_{\mathrm c}^3/(2k_{\mathrm J}^2k_{\mathrm q})$ (Eq. \eqref{eq:cutoff_exact}), the numerator is further simplified to
\begin{equation}
\left(\frac{\partial D(\gamma,k)}{\partial k}\right)_{k=k_{\mathrm c}} = \frac{3}{k_{\mathrm c}}-\frac{k_{\mathrm J}^2}{k_{\mathrm c}^3}\left(1 - \frac{k_{\mathrm c}^4}{2k_{\mathrm J}^2k_{\mathrm q}^2}\right).
\end{equation}

Next we derive the denominator as
\begin{align}
\frac{\partial D(\gamma,k)}{\partial \gamma} &= \frac{\partial \zeta}{\partial \gamma}\frac{\partial D(\zeta,k)}{\partial \zeta}\nonumber \\
 &= -i\frac{k_{\mathrm J}^2k_{\mathrm q}}{2k^4v_\mathrm{t}} \left(Z^\prime \left(\zeta-\frac{k}{2k_{\mathrm q}} \right)-Z^\prime \left(\zeta+\frac{k}{2k_{\mathrm q}} \right) \right) \nonumber \\
 &= i\frac{k_{\mathrm J}^2k_{\mathrm q}}{k^4v_\mathrm{t}} \left(\left( \zeta-\frac{k}{2k_{\mathrm q}}\right)Z\left( \zeta-\frac{k}{2k_{\mathrm q}}\right)-\left( \zeta+\frac{k}{2k_{\mathrm q}}\right)Z\left( \zeta+\frac{k}{2k_{\mathrm q}}\right)\right).
\end{align}
Similarly to the numerator, it is simplified for $k_{\mathrm c}$ as
\begin{equation}
\left(\frac{\partial D(\gamma,k)}{\partial \gamma}\right)_{k=k_{\mathrm c}} = \sqrt{\pi}\frac{k_{\mathrm J}^2}{k_{\mathrm c}^3v_\mathrm{t}}\exp\left(-\frac{k_{\mathrm c}^2}{4k_{\mathrm q}^2} \right).
\end{equation}

Finally, we obtain an explicit form of $\left(\partial \gamma/\partial k\right)_{k=k_{\mathrm c}}$ as
\begin{equation}
\left(\frac{\partial \gamma}{\partial k}\right)_{k=k_{\mathrm c}} = -\frac{\frac{3}{k_{\mathrm c}}-\frac{k_{\mathrm J}^2}{k_{\mathrm c}^3}\left(1 - \frac{k_{\mathrm c}^4}{2k_{\mathrm J}^2k_{\mathrm q}^2}\right)}{\sqrt{\pi}\frac{k_{\mathrm J}^2}{k_{\mathrm c}^3v_\mathrm{t}}\exp\left(-\frac{k_{\mathrm c}^2}{4k_{\mathrm q}^2} \right)} = -\frac{v_\mathrm{t}}{\sqrt{\pi}}\left( \frac{3k_{\mathrm c}^2}{k_{\mathrm J}^2}+\frac{k_{\mathrm c}^4}{2k_{\mathrm J}^2k_{\mathrm q}^2}-1 \right)\exp\left(\frac{k_{\mathrm c}^2}{4k_{\mathrm q}^2} \right).
\label{eq:cutoff_slope}
\end{equation}

Let us see how this slope asymptotes in the classical limit. In the limit of $\hbar \rightarrow 0$, $k_{\mathrm c}=k_{\mathrm J}$ and $k_{\mathrm q} = \infty$, and it is straightforward to show
\begin{equation}
\left(\frac{\partial \gamma}{\partial k}\right)_{k=k_{\mathrm c}} =-\frac{2v_\mathrm{t}}{\sqrt{\pi}}.
\end{equation}

We first solve Eq. \eqref{eq:cutoff_exact} for $k_{\mathrm c}$ numerically, and obtain the slope from Eq. \eqref{eq:cutoff_slope} for various values of $\alpha$. Figure \ref{fig:kcutoff-alpha} (b) shows the cutoff slope derived from Eq. \eqref{eq:cutoff_slope} across different $\alpha$ values. The negative sign is applied in the figure for better visual clarity. The slope magnitude shows a smooth decrease starting from larger $\alpha$, reaches a minimum value around $\alpha\sim 1$, and then increases steeply for $\alpha \lesssim 0.5$. In contrast to the slight variation of the cutoff wavenumber within this range of $\alpha$, the spectral slope exhibits a sharp crossover.

The rapid increase in the slope magnitude for small values of $\alpha$ arises from the denominator in Eq. \eqref{eq:cutoff_slope}. This denominator is proportional to $\exp(-k_{\mathrm c}^2/4k_{\mathrm q}^2) = \exp(-v_\mathrm{recoil}^2/v_\mathrm{t}^2)$, where $v_\mathrm{recoil} = \hbar k_{\mathrm c}/(2m)$ represents half the recoil velocity of a particle for a wave with $k = k_{\mathrm c}$ \citep{mendoncaLandauDampingParticle2023}. Thus, the denominator characterizes the population of particles at half the recoil speed. As $\alpha$ decreases, the recoil speed shifts toward the exponential tail of the velocity distribution function, leading to a rapid decline in the number of particles. This depletion of free-streaming particles, rather than a change in the cutoff wavenumber, serves as the physical origin of the sharp variation in the spectral slope around $\alpha \sim 0.5$.

Here we define a critical $\alpha_c$ that characterizes the crossover between quantum-dominant and thermal-dominant systems. This critical value corresponds to the cutoff wavenumber reaching its maximum value at $\alpha \sim 0.5$ as shown in Figure \ref{fig:kcutoff-alpha} (a). We theoretically derive this critical value from
\begin{equation}
\frac{\partial k_{\mathrm c}}{\partial \alpha} =
-\frac{
\frac{\partial\left(F\left(\frac{k_{\mathrm c}}{2k_{\mathrm q}}\right)-\frac{k_{\mathrm c}^3}{2k_{\mathrm J}^2 k_{\mathrm q}}\right)}{\partial \alpha}
}
{
\frac{\partial\left(F\left(\frac{k_{\mathrm c}}{2k_{\mathrm q}}\right)-\frac{k_{\mathrm c}^3}{2k_{\mathrm J}^2 k_{\mathrm q}}\right)}{\partial k_{\mathrm c}}
}
=0,
\end{equation}
which allows us to focus only on the numerator as
\begin{equation}
\frac{\partial\left(F\left(\frac{k_{\mathrm c}}{2k_{\mathrm q}}\right)-\frac{k_{\mathrm c}^3}{2k_{\mathrm J}^2 k_{\mathrm q}}\right)}{\partial \alpha} =0.
\label{eq:eq_critical_alpha}
\end{equation}
Eq. \eqref{eq:eq_critical_alpha} finally leads to
\begin{equation}
F\left( \frac{k^\prime}{\sqrt{2}\alpha_c} \right) = \frac{k^\prime}{\sqrt{2}\alpha_c\left(\frac{k^{\prime 2}}{\alpha_c^2}-1\right)},
\label{eq:eq_critical_alpha_norm2}
\end{equation}
where $k^\prime = k_{\mathrm c}/k_{\mathrm{qJ}}$. If we write $\beta = k^\prime/(\sqrt{2}\alpha_c)$, Eq. \eqref{eq:eq_critical_alpha_norm2} simplifies to
\begin{equation}
F\left(\beta \right) = \frac{\beta}{2\beta^2-1},
\end{equation}
which can be solved numerically for $\beta$. Once we find $\beta$, $\alpha_c$ is obtained from Eq. \eqref{eq:cutoff_exact} in the form of 
\begin{equation}
\alpha_c = \left(\frac{F(\beta)}{2\beta^3}\right)^{\frac{1}{4}}=\left(\frac{1}{2\beta^2\left(2\beta^2-1\right)}\right)^{\frac{1}{4}}.
\end{equation}
From the numerical analysis, we find $\beta=1.501975...$, and consequently obtain the critical quantum-to-thermal ratio as
\begin{equation}
\alpha_c = 0.501218... \approx 0.5.    
\end{equation}

\section{Implications for observations}
\label{sec:implication}
\begin{figure}
\includegraphics[width=1.0\linewidth]{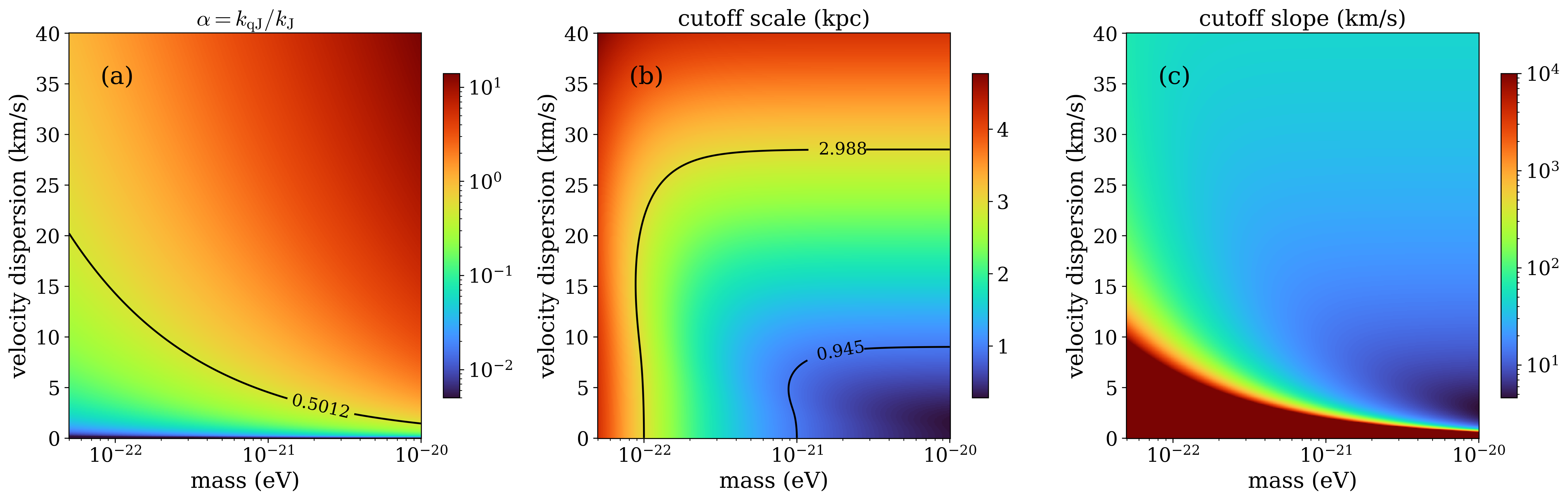}
\caption{(a) $\alpha$, (b) the cutoff scale of $2\pi/k_{\mathrm c}$ (kiloparsec), and (c) the absolute value of the cutoff slope $\mathrm{(km~s^{-1})}$ for parameter ranges of FDM mass (eV) and velocity dispersion $\mathrm{(km~s^{-1})}$. The color scale is saturated at $10,000~{\mathrm{km~s^{-1}}}$ in (c).}
\label{fig:cutoff-physics-scale}
\end{figure}
In this section, we quantify the characteristic parameters $\alpha$, $k_{\mathrm c}$, and $(\partial \gamma/\partial k)_{k=k_{\mathrm c}}$ using the physical constants of $G = 6.67\times10^{-8}\ \mathrm{cm^3\ g^{-1}\ s^{-2}}$ and $\hbar = 1.05\times10^{-27}\ \mathrm{erg\ s}$, with a background mass density $\rho_0 = 2.25\times10^{-24}\ \mathrm{g\ cm^{-3}}$. This background density is obtained from the present-day dark matter density by applying the scaling factor $(1+z)^3$ at a redshift of $z = 99$ with a CDM density of $\Omega_{\mathrm{c}} h^2 \simeq 0.12$ \citep{aghanimPlanck2018Results2020}, where $\Omega_{\mathrm{c}}$ is the CDM density parameter and $h$ is the dimensionless Hubble parameter. The redshift of $z=99$ corresponds to the epoch at which the linear instability grows. The characteristic parameters are then expressed in terms of the FDM mass $m$ and the velocity dispersion $v_\mathrm{t}$, as illustrated in Figure \ref{fig:cutoff-physics-scale}.

Figure \ref{fig:cutoff-physics-scale} (b) shows the cutoff scale in kiloparsecs (kpc). If we adopt a characteristic scale of 3.0 kpc as the quantum Jeans scale for a mass of $10^{-22}~\mathrm{eV}$ (the case for a mass of $10^{-21}~\mathrm{eV}$ is also shown), this same cutoff scale can be realized by a range of velocity dispersions, reaching up to $30~\mathrm{km~s^{-1}}$, as indicated by the contour line in the panel. Along the 3.0 kpc contour, the system transitions from a quasi-QHD regime to one dominated by thermal effects through an intermediate state, as also indicated by the $\alpha$ profile in Figure \ref{fig:cutoff-physics-scale} (a). Consequently, the cutoff slope magnitude varies significantly across the contour level $\alpha=\alpha_c$, as shown in Figures \ref{fig:cutoff-physics-scale} (a) and \ref{fig:cutoff-physics-scale} (c).

Lyman-alpha forest data provide valuable information about small-scale cosmic structures \citep{irsicFirstConstraintsFuzzy2017,armengaudConstrainingMassLight2017}, whose suppression is encoded in the transfer function that shapes the linear matter power spectrum of FDM \citep{huFuzzyColdDarkMatter2000}. Our quantum-kinetic linear analysis determines the spectral shape, thereby characterizing the small-scale form of this transfer function. The resulting matter power spectrum can then be compared with observations. Conversely, once the cutoff scale and spectral shape are measured, a transfer function incorporating our linear analysis could, in principle, allow us to infer the mass and velocity dispersion of FDM simultaneously. A detailed quantitative comparison is left to future work.

\section{Conclusion}
\label{sec:conclusion}
In this paper, we have presented a kinetic theory of the gravitational instability that fully incorporates quantum effects. The theory enables us to quantify not only the cutoff wavenumber but also the spectral slope at the cutoff, thereby providing an accurate description of the collisionless damping of the instability and insight into the behavior of fuzzy dark matter on the smallest scales in the universe. We found that the quantum-thermal ratio, defined as $\alpha = k_{\mathrm{qJ}}/k_{\mathrm J}$, controls the properties of the instability and the overall spectral shape. While the cutoff scale remains close to the quantum Jeans scale as long as $\alpha \lesssim 0.76$, the spectral slope near the cutoff changes significantly across $\alpha \sim 0.5$. These characteristics suggest that the cutoff scales and spectral shapes of observable matter power spectra, imprinted in Lyman-alpha forest data, could be key to constraining the mass and velocity dispersion of fuzzy dark matter.

Nonlinear numerical simulations are expected to reveal the subsequent development of the collapsed density structure and its profile at galactic scales, particularly in relation to the core-cusp problem. Even if the instability arises from the thermally dominant (incoherent) state, the bosonic particles would transition to the ground (coherent) state inside the gravitationally collapsed region. Since the density increases significantly beyond its initial level, the gravitational interaction timescale shortens, allowing the particles to transition into a BEC within the age of the universe \citep{levkovGravitationalBoseEinsteinCondensation2018}.

However, current standard numerical simulations, which are based on the Schr\"odinger-Poisson equations with a single wave function, are insufficient to clarify this transition. A more suitable approach would be to solve the Wigner transport equation (Eq. \eqref{eq:wigner_transport_fourier}) by adapting Vlasov simulation algorithms (e.g., \citep{yoshikawa400TrilliongridVlasov2021}). Although tracking the evolution of the Bose-Einstein distribution function with a limited number of numerical cells in momentum space poses computational challenges, this approach would enable a precise, self-consistent treatment of the evolution of fuzzy dark matter. This, in turn, would help elucidate the formation of soliton cores at galactic scales and the creation of boson stars \citep{levkovGravitationalBoseEinsteinCondensation2018}.

\begin{acknowledgments}
This work was supported by JSPS KAKENHI Grant Number JP25H00625.
\end{acknowledgments}

\appendix
\section{Semi-classical expansion of the Wigner transport equation}
\label{sec:appendix_semiclassical}
The Wigner transport equation (Eq.~\eqref{eq:wigner_transport_fourier}) is equivalently expressed in the differential form \citep{moyalQuantumMechanicsStatistical1949} as
\begin{equation}
  \frac{\partial f_\mathrm{w}}{\partial t} = \frac{2}{\hbar}H\sin\!\left(\frac{\hbar}{2}
  \left(\overleftarrow{\partial_x}\overrightarrow{\partial_p}
       -\overleftarrow{\partial_p}\overrightarrow{\partial_x}\right)\right)f_\mathrm{w},
\end{equation}
where $\overleftarrow{\partial}$ and $\overrightarrow{\partial}$ operate on the left- and right-hand sides, respectively. Expanding $\sin(x) \approx x - \frac{x^3}{6} + \cdots$, we obtain the
semi-classical Wigner--Vlasov equation \citep{bertrandClassicalVlasovPlasma1980,koppSolvingVlasovEquation2017,moczSchrodingerPoissonVlasovPoissonCorrespondence2018,gomesQuantumKineticTheory2023}
\begin{equation}
  \frac{\partial f_\mathrm{w}}{\partial t} = \{H,f_\mathrm{w}\}
  - \frac{\hbar^2}{24}\left(
      \frac{\partial^3 H}{\partial x^3}\frac{\partial^3 f_\mathrm{w}}{\partial p^3}
     -\frac{\partial^3 H}{\partial p^3}\frac{\partial^3 f_\mathrm{w}}{\partial x^3}
    \right),
   \label{eq:semi-classical}
\end{equation}
where $\{H,f_\mathrm{w}\} = \partial_x H\ \partial_p f_\mathrm{w} - \partial_p H\ \partial_x f_\mathrm{w}$ is the classical Poisson bracket.

\section{Plasma dispersion function}
\label{sec:appendix_plasma}
The plasma dispersion function \citep{friedPlasmaDispersionFunction1961,xieRapidComputationPlasma2024} is defined for $\Im(\zeta)>0$, and is analytically continued as
\begin{equation}
  Z(\zeta) = 
  \begin{cases}
    \frac{1}{\sqrt{\pi}}\int_{-\infty}^{\infty}\frac{\exp{(-z^2)}}{z-\zeta}\ dz & (\Im(\zeta) > 0), \\
    \frac{1}{\sqrt{\pi}}\mathcal{P}\int_{-\infty}^{\infty}\frac{\exp{(-z^2)}}{z-\zeta}\ dz+i\sqrt{\pi}\exp{(-\zeta^2)} & (\Im(\zeta) = 0), \\
    \frac{1}{\sqrt{\pi}}\int_{-\infty}^{\infty}\frac{\exp{(-z^2)}}{z-\zeta}\ dz+ 2i\sqrt{\pi}\exp{(-\zeta^2)} & (\Im(\zeta) < 0),
  \end{cases}
  \label{eq:Z_full}
\end{equation}
where $\mathcal{P}\int$ denotes the Cauchy principal value. The principal value is also related to
\begin{equation}
\frac{1}{\sqrt\pi}\mathcal{P}\int_{-\infty}^{\infty} \frac{\exp{(-z^2)}}{z-\zeta}dz = -2F(\zeta),
\end{equation}
where
\begin{equation}
    F(u) \equiv \exp{(-u^2)}\!\int_0^u \exp{(t^2)}dt
\end{equation}
is the Dawson integral.

We straightforwardly obtain the first derivative of the plasma dispersion function as
\begin{equation}
  Z^\prime(\zeta) = -2\left(1+\zeta Z(\zeta)\right).
  \label{eq:Zprime}
\end{equation}

\bibliography{qkpaper2026}

\end{document}